\documentclass{aa}  

\usepackage{ulem}
\usepackage{color}
\definecolor{orange}{rgb}{1, 0.5, 0.}
\definecolor{pink}{rgb}{0.7, 0.5, 0.5}
\newcommand{\dele}[1]{} %accepted CR comment
\usepackage{array}

\usepackage{graphicx}
\usepackage{txfonts}
\usepackage[colorlinks=true,citecolor=blue]{hyperref}
\usepackage{siunitx}
\usepackage{xspace}
\usepackage{mathtools}

\defcitealias{2020Sci...369.1347Meneghetti}{M20}
\defcitealias{2014MNRAS.438..195Planelles}{P14}
\defcitealias{2015ApJ...813L..17Rasia}{R15}
\defcitealias{2020A&A...642A..37Bassini}{B20}
\defcitealias{2018MNRAS.479.1125RagoneFigueroa}{RF18}

\newcommand{\appropto}{\mathrel{\vcentre{
  \offinterlineskip\halign{\hfil$##$\dCR
    \propto\dCR\noalign{\kern2pt}\sim\dCR\noalign{\kern-2pt}}}}}

\titlerunning{$V_{\rm max}\left(M_{\rm SH}, \epsilon\right)$}
\authorrunning{Ragagnin et al.}
\begin{document}

   \title{Galaxies in the central regions of simulated galaxy clusters}

   %\subtitle{I. Overviewing the $\kappa$-mechanism}

   \author{Antonio Ragagnin\inst{\ref{unibo},\ref{oats},\ref{ifpu}}\thanks{\email{antonio.ragagnin@unibo.it}},
   Massimo Meneghetti\inst{\ref{oas},\ref{infnbo}},
   Luigi Bassini\inst{\ref{ctac}},
   Cinthia Ragone-Figueroa\inst{\ref{iateconicet},\ref{oats}},
   Gian Luigi Granato\inst{\ref{oats},\ref{iateconicet},\ref{ifpu}},
   Giulia Despali\inst{\ref{heidelberg}},
   Carlo Giocoli\inst{\ref{oas},\ref{infnbo},\ref{unibo}},
    Giovanni Granata\inst{\ref{unimi}},
    Lauro Moscardini\inst{\ref{unibo},\ref{oas},\ref{infnbo}}, 
    Pietro Bergamini\inst{\ref{unimi},\ref{oas}}, 
    Elena Rasia\inst{\ref{oats}, \ref{ifpu}}, 
    Milena Valentini\inst{\ref{usm}},
    Stefano Borgani\inst{\ref{units},\ref{oats}},
    Francesco Calura\inst{\ref{oas}},
    Klaus Dolag\inst{\ref{usm},\ref{mpa}},
    Claudio Grillo\inst{\ref{unimi},\ref{oami}},
    Amata Mercurio\inst{\ref{oana}}, 
    Giuseppe Murante\inst{\ref{oats}, \ref{ifpu}},  
    Priyamvada  Natarajan\inst{\ref{yale}}, 
    Piero Rosati\inst{\ref{oami},\ref{unife}}, 
    Giuliano Taffoni\inst{\ref{oats}},
    Luca Tornatore\inst{\ref{oats}}, 
    Luca Tortorelli\inst{\ref{usm}}
   }
   \institute{
    Dipartimento di Fisica e Astronomia "Augusto Righi", Alma Mater Studiorum Università di Bologna, via Gobetti 93/2, I-40129 Bologna, Italy\label{unibo}
            \and
            INAF - Osservatorio Astronomico di Trieste, via G.B. Tiepolo 11, I-34143 Trieste, Italy\label{oats}
            \and
            IFPU - Institute for Fundamental Physics of the Universe, Via Beirut 2, I-34014 Trieste, Italy\label{ifpu}
            \and
                        INAF-Osservatorio di Astrofisica e Scienza dello Spazio di Bologna,
Via Piero Gobetti 93/3, I-40129 Bologna, Italy\label{oas}
            \and
            INFN-Sezione di Bologna, Viale Berti Pichat 6/2, I-40127 Bologna,
Italy\label{infnbo}
            \and
            centre for Theoretical Astrophysics and Cosmology, Institute for Computational Science, University of Zurich, Winterthurerstrasse 190, CH-8057 Zürich, Switzerland\label{ctac}
            \and
            Instituto de Astronom\'ia Te\'orica y Experimental (IATE), Consejo Nacional de Investigaciones Cient\'ificas y T\'ecnicas de la\\ Rep\'ublica Argentina (CONICET), Universidad Nacional de C\'ordoba, Laprida 854, X5000BGR, C\'ordoba, Argentina
            \label{iateconicet}
            \and Zentrum für Astronomie der Universität Heidelberg, Institut für Theoretische Astrophysik, Albert-Ueberle-Str. 2, D-69120 Heidelberg, Germany\label{heidelberg}
            \and Dipartimento di Fisica, Università degli Studi di Milano, Via Celoria 16, I-20133 Milano, Italy\label{unimi}
            \and
            Universitäts-Sternwarte, Fakultät für Physik, Ludwig-Maximilians-Universität München, Scheinerstr.1, 81679 München, Germany \label{usm}%\\
            \and
            Astronomy Unit, Department of Physics, University of Trieste, via Tiepolo 11, I-34131 Trieste, Italy \label{units}%\\
            \and
            Max-Planck-Institut f\"{u}r  Astrophysik (MPA), Karl-Schwarzschild Strasse 1, D-85748 Garching bei M\"{u}nchen, Germany\label{mpa}%\\
            \and INAF - IASF Milano, via A. Corti 12, I-20133 Milano, Italy\label{oami}
            \and INAF-Osservatorio Astronomico di Capodimonte, Via Moiariello
16, 80131 Napoli, Italy\label{oana}
            \and Department of Astronomy, Yale University, New Haven, CT, USA\label{yale}
            \and Dipartimento di Fisica e Scienze della Terra, Università degli Studi di Ferrara, Via Saragat 1, I-44122 Ferrara, Italy\label{unife}
           % \and Institute for Particle Physics and Astrophysics, ETH Zürich, Wolfgang-Pauli-Str. 27, 8093
Zürich, Switzerland\label{ippa}
            }

   \date{submitted}

% \abstract{}{}{}{}{} 
% 5 {} token are mandatory
 
  \abstract
  % context heading (optional)
  {\cite{2020Sci...369.1347Meneghetti} (hereafter M20) found that observed cluster member galaxies are more compact than their counterparts in $\Lambda$CDM hydrodynamic simulations, as indicated by the difference in their  strong gravitational lensing properties. \citetalias{2020Sci...369.1347Meneghetti} reported that the measured and simulated galaxy-galaxy strong lensing  events on small scales are discrepant by one order of magnitude. Among possible explanations for this discrepancy, some studies suggested that simulations with better resolution and implementing different schemes for galaxy formation, compared to the ones used in \citetalias{2020Sci...369.1347Meneghetti}, could bring simulations in better agreement with the observations. 
  }
  % aims heading (mandatory)
   {In this paper, we assess the impact of numerical resolution and of the implementation of energy input from AGN feedback models on the inner structure of cluster sub-haloes in hydrodynamic simulations.}
  % methods heading (mandatory)
   {We compare several zoom-in re-simulations of a sub-sample of the cluster-sized haloes studied in \citetalias{2020Sci...369.1347Meneghetti}, obtained  by varying mass resolution,  softening length and AGN energy feedback scheme. We study the impact of these different setups on the subhalo (SH) abundances, their radial distribution, their  density and mass profiles and the relation between the maximum circular velocity, which is a proxy for SH compactness. 
   }
   {Regardless of the adopted numerical resolution and feedback model, SHs with masses $M_{\rm SH}\lesssim 10^{11}\;h^{-1}\;M_\odot$, the most relevant mass-range for galaxy-galaxy strong lensing, have maximum circular velocities $\sim 30\%$ smaller than those measured from strong lensing observations of \cite{2019A&A...631A.130Bergamini}. We also find that simulations with  less effective AGN energy feedback produce massive SHs ($M_{\rm SH} \gtrsim 10^{11}\;h^{-1}\;M_\odot$) with higher maximum circular velocity and that their $V_{\rm max}-M_{\rm SH}$ relation approaches the observed one. However the stellar-mass number count of these objects exceeds the one found in observations and we find that the compactness of these simulated SHs  is the result of an extremely over-efficient star formation in their cores, also leading to larger-than-observed SH stellar mass. 
   }
  % conclusions heading (optional), leave it empty if necessary 
   {Regardless of the resolution and galaxy formation model adopted, simulations are unable to simultaneously reproduce the observed stellar masses and compactness (or maximum circular velocities) of cluster galaxies. Thus, the discrepancy between theory and observations that emerged from the analysis of \citetalias{2020Sci...369.1347Meneghetti} persists. It remains an open question as to whether such a discrepancy reflects limitations of the current implementation of galaxy formation models or the $\Lambda$CDM paradigm.}

  % \keywords{Galaxies: clusters: general -- Cosmology: cosmological parameters -- Galaxy: formation -- method: numerical -- Hydrodynamics }

   \maketitle

%-------------------------------------------------------------------

\section{Introduction}
\label{sec:intro}
%\cgiocoli{Eviterei gli acronimi GC, SH cosi' nell'introduzione. Rendono difficile la lettura, secondo me. Per questioni generali userei M per la massa dell'alone, m per la massa della sottostruttura.}
The properties and abundances of cluster sub-haloes and their associated galaxies are important probes of cosmology \citep[see e.g.][]{PNKneib1997,1998ApJ...499L...5Moore,1998MNRAS.299..728Tormen,PNSpringel2004,2004MNRAS.355..819Gao,2005MNRAS.356.1233VanDenBosch,2012ARA&A..50..353Kravtsov,2016MNRAS.456.2486Despali,2017MNRAS.469.1997Despali,2019BAAS...51c..73Natarajan,2021arXiv211005498Ragagnin}, and galaxy formation ~\citep[][]{2015MNRAS.448.1835Taylor}. In particular, they can be used to test the predictions of the Cold Dark Matter (CDM) paradigm  of structure formation on multiple scales in clusters ~\citep[see e.g.][]{2007MNRAS.376..180Natarajan,2021arXiv210202375Yang}.
%, and mock observations~\citep{2012MNRAS.421.3343Giocoli}. 

Gravitational lensing has proved to be a powerful tool to map the distribution of  dark matter (DM) in galaxy clusters both on large and small scales~\citep[as in][]{Grillo_2015,2015ApJ...801...44Zitrin,2017MNRAS.472.3177Meneghetti}  as shown in reviews by \cite{KneibPN2011,2020A&ARv..28....7Umetsu}. 
In particular, recent improvements in combined strong plus weak lensing modelling techniques combined with galaxy kinematic measurements from integral field spectroscopy, enabled constraining the matter distribution in cluster substructures in great detail \citep[see e.g.][]{PN+2019,2019A&A...631A.130Bergamini}.

Based on cluster reconstructions, \cite{2020Sci...369.1347Meneghetti} (M20) recently reported that lensing clusters observed in the CLASH survey \citep{2012ApJS..199...25Postman} and Frontier Fields \citep{2017ApJ...837...97Lotz} HST programmes have cross sections for galaxy-galaxy strong projected  lensing (GGSL) that exceed by one order of magnitude the expectations in the context of current $\Lambda$CDM-based hydrodynamic simulations. The ability of subhaloes (SHs) to produce strong lensing events depends primarily on their compactness and location within the cluster. More compact SHs lying at smaller projected cluster-centric radii can more easily exceed the critical surface mass density required for strong lensing and can therefore be more effective at splitting and strongly distorting the images of background galaxies. 

The spatial distribution of cluster SHs can be traced by the galaxies that they host. The maximum circular velocity, defined as 
\begin{equation}
    V_{\rm max}=\mathrm{max} \left( \sqrt{\frac{GM(<r)}{r}} \right)\;, 
    \label{eq:vcirc}
\end{equation}
where $r$ is the three-dimensional distance from the SH centre, $M(<r)$ is SH radial mass profile, and $G$ is the gravitational constant. 
The value of $V_{\rm max}$ has been demonstrated to be a robust proxy for SH compactness. As shown by \cite{2019A&A...631A.130Bergamini}, the SH circular velocity can be measured by combining imaging and spectroscopic observations of strong lensing. 

\citetalias{2020Sci...369.1347Meneghetti} noted that the galaxies with a projected distance within $\sim 0.15 R_{\rm vir}$ (where $R_{\rm vir}$ is the cluster virial radius\footnote{We denote as $R_\Delta$ and $M_\Delta$ the radius and mass of the sphere enclosing a density equal to $\Delta$ times the critical density at the respective redshift.  See \cite{2015MNRAS.447.1873Naderi} for a review on galaxy cluster masses and radii definitions.}) in the clusters reconstructed by  \cite{2019A&A...631A.130Bergamini} have maximum circular velocities exceeding  those of SHs with the same mass in the hydrodynamic simulations of \cite{2015ApJ...813L..17Rasia} (hereafter R15) by $\sim 30-50\%$ on average. They also found that the number density of these SHs near the cluster critical lines in observations is higher than in simulations. Thus, \citetalias{2020Sci...369.1347Meneghetti} concluded that the larger GGSL cross section reflects the more compact spatial distribution and internal structure of observed SHs compared to simulated ones. 
 
This mismatch  may arise due to limitations in our current simulations or may warrant revisiting our assumptions on the nature of dark matter~\citep[see e.g.][]{2020MNRAS.491.1295Despali,2021arXiv210608292Bhattacharyya,2021arXiv210202375Yang,2021MPLA...3630001Nguyen}. Moreover, it has been argued that the simulation results might depend on mass resolution and artificial tidal disruption, that could impact the properties of SHs~\citep{2021MNRAS.503.4075Green}. \citetalias{2020Sci...369.1347Meneghetti} performed a comparison between low- and high-resolution re-simulations of a single cluster and did not notice significant differences between their derived GGSL cross sections. Nevertheless, in a recent paper, \cite{2021MNRAS.505.1458Bahe} claimed that cluster SHs in the very high-resolution Hydrangea simulation suite are in better agreement with observations \citep[see also][]{2021MNRAS.504L...7Robertson}. 
We will devote part of this paper to assessing the origin of these two contradicting claims.

On top of this, most high-resolution hydrodynamic simulations have problems in reproducing correct stellar masses of high mass SHs.
For instance, \citet[][RF18]{2018MNRAS.479.1125RagoneFigueroa}, showed that
the stellar masses of the observed  brightest cluster galaxies (BCGs)
are clearly overproduced by the Hydrangea and IllustrisTNG simulations (see their Fig. 1). The excess of star formation in these simulations \citep[for instance the one shown in Fig. 3 in][on the IllustrisTNG simulations]{2014MNRAS.445..175Genel} is likely related to a general difficulty in modelling AGN feedback at the high stellar mass regime. 

In other words, it seems that these simulations are not able to match observations of stellar masses  (and therefore their baryon fractions), measured luminosities, internal structure and lensing properties simultaneously.

It is important to note here that in addition to numerical resolution, the inner structure of SHs is very sensitive to the details of   sub-grid models for cooling, star formation, and feedback implemented in the simulations.
As \citetalias{2020Sci...369.1347Meneghetti} suggested, numerical effects may also play an important role in affecting the  discrepancy they report. \citetalias{2018MNRAS.479.1125RagoneFigueroa} pointed out that particular care must be taken in controlling the centring of super-massive black hole (SMBH) particles, which leave their host galaxies in the course of the simulations due to frequent dynamical disturbances and mergers, whose effects are probably amplified by numerical limitations. Under these circumstances, the SMBH energy feedback does not effectively suppress gas over-cooling~\citep{2006MNRAS.367.1641Borgani,2013MNRAS.431.2513Wurster} nor star formation in the centre of massive galaxies \citep{2020MNRAS.493.2926Ludlow,2021arXiv210901489Bahe}. %Increasing the resolution makes it even more challenging to keep these multiple effects under control.
Thus it is worth studying the impact of feedback schemes, softening and resolution in producing the $M_{\rm SH}-V_{\rm max}$ relation.
 
Moreover, when comparing density profiles of simulated haloes to observations, it is crucial to take into account that different softening lengths $\epsilon$  lead to varying minimum well-resolved radii, below which density profiles cannot be trusted as the gravitational force gets under-estimated\footnote{For two particles with distance $r$, their gravitational potential is computed as  $\Phi\propto1/\sqrt{r^2+\epsilon^2}$}.
This minimum radius for density profiles is often estimated as $2.8$ times the 
fiducial equivalent softening, i.e. $h\approx 2.8\epsilon$ \citep[see ][for more details on the choice of this value]{1987ApJS...64..715Hernquist}.
For these reasons, in order to disentangle the effect of resolution alone, in this work we will use simulations with varying softening, resolution and feedback parameters %\cgiocoli{
in order to tackle all these small-scale issues.
%}. 

The paper is organised as follows. In Sec. \ref{sec:sims} we present our simulations in detail. In Sec.  \ref{sec:discu}, we discuss the results in terms of   mass-velocity SH scaling relations and present our conclusions and discussion thereof in Sec. \ref{sec:conclu}.

\section{Numerical Setup}
\label{sec:sims}

\begin{table}
\caption{List of feedback and resolution parameters  for the four suites 1xR15, 1xRF18, 10xB20 and its 1x counterpart (1xB20).  Rows show respectively: minimum gas temperature $T_g,$  outflow efficiency $\epsilon_o$, BH radiative efficiency $\epsilon_r,$ BH feedback factor $\epsilon_f,$ dark matter softening $\epsilon_{\rm DM},$  stellar softening  $\epsilon_\star,$ and dark matter particle mass $m_{\rm DM}.$}    
% title of Table
\label{tbl:ps}      % is used to refer this table in the text
\centering                                      % used for centreing table
\begin{tabular}{r |r| r| r r} 
\hline\hline
          & 1xR15 & 1xRF18 & 10xB20  & 1xB20\\
          \hline
min $T_g [{\rm K}]$   &   $50$    &   $50$ &        \multicolumn{2}{c}{$20$}\\
$\epsilon_o$  & $0.15$ &   $-$ &     \multicolumn{2}{c}{$-$ }  \\
$\epsilon_r$  & $0.1$ &   $0.07$ &     \multicolumn{2}{c}{$0.07$ }  \\
$\epsilon_f$ & $0.05$ &  $0.1$ &   \multicolumn{2}{c}{$0.16$} \\
 $\epsilon_{\rm DM}  [h^{-1\;}a{\rm kpc}]$   &  $3.75$ & $5.62$  & $1.4$ & $3.0$  \\
 $\epsilon_\star  [h^{-1\;}a{\rm kpc}]$     & $2.0$   & $3.0$     & $0.35$  & $0.75$ \\
 $m_{\rm DM}[10^8\;h^{-1} {\rm M}_\odot]$  & $8.3$ & $8.3$  & $0.83$ & $8.3$\\
Reference & \protect\citetalias{2015ApJ...813L..17Rasia} &  \protect\citetalias{2018MNRAS.479.1125RagoneFigueroa} & \multicolumn{2}{c}{{\protect\citetalias{2020A&A...642A..37Bassini}}}\\
\end{tabular}
\end{table}

\begin{table}
\caption{Cluster haloes studied in this work. First column reports the name of the $6$ out of the $29$ Dianoga regions used in this work. The panels in other columns are the halo virial masses ($M_\mathrm{vir}$) and the number  $N_\mathrm{s}$ of SHs with mass $M_\mathrm{\rm SH}>2\times10^{10}h^{-1}{\rm M}_\odot$ within a distance of $r<0.15R_\mathrm{vir}$ from the halo centre. These quantities are reported for all suites (1xR15, 1xRF18, 10x, and 1xB20). Virial masses
are in units of $10^{14}h^{-1}{\rm M}_\odot.$ Haloes are extracted from a the output at redshift $z=0.4.$
}
\label{tbl:ds}      % is used to refer this table in the text
\centering                                      % used for centreing table
\begin{tabular}{r |r r |  r r | r r| r r} 
\hline\hline
Name & \multicolumn{2}{c}{1xR15} &  \multicolumn{2}{c}{1xRF18} & \multicolumn{2}{c}{10xB20}  & \multicolumn{2}{c}{1xB20} \\
\hline
     & $M_\mathrm{vir}  $ & $N_\mathrm{s}$  & $M_\mathrm{vir}  $ & $N_\mathrm{s}$ & $M_\mathrm{vir}  $ & $N_\mathrm{s}$  & $M_\mathrm{vir}  $ & $N_\mathrm{s}$\\
%     & $[10^{14}{\rm M}_\odot]$ &   & $[10^{14}{\rm M}_\odot]$ &    & $[10^{14}{\rm M}_\odot]$ &\\   
     \hline
D1  & 11.4 & 62  & 11.4 & 38  & 11.1 & 87  \\
D3  & 4.9 & 25  & 4.8 & 13  & 4.8 & 28  & 4.8 & 21  \\
D6  & 6.3 & 28  & 6.3 & 25  & 6.4 & 42  & 6.4 & 34  \\
D10  & 10.4 & 70  & 10.4 & 48  & 10.1 & 83  \\
D18  & 8.0 & 32  & 8.1 & 23  & 7.7 & 49  & 7.6 & 31  \\
D25  & 3.1 & 13  & 3.0 & 12  & 3.0 & 22  & 2.3 & 13  \\
\end{tabular}
\end{table}

The Dianoga suite of simulations consists of a set of $29$ regions containing cluster-size haloes extracted from a parent DM-only cosmological box of side-length $1$ comoving Gpc$/h.$
These regions were re-simulated including baryons using the zoom-in initial conditions from \cite{2011MNRAS.418.2234Bonafede} and assuming different 
baryonic physics models, softening, and mass resolutions. 

The simulations are performed using the code P-Gadget3 \citep{2001MNRAS.328..726Springel}, adopting an improved Smoothed Particle Hydrodynamics (SPH)  solver \citep{2016MNRAS.455.2110Beck}, and a stellar evolution scheme \citep{2007MNRAS.382.1050Tornatore,2005Natur.435..629Springel}   which follows  11 chemical  elements (H, He,
C, N, O, Ne, Mg, Si, S, Ca, Fe) with the aid of the CLOUDY photo-ionisation
code \citep{1998PASP..110..761Ferland}. 
A general description of how the SMBH  and energy feedback are modelled can be found in  \cite{2005Natur.435..629Springel,2010MNRAS.401.1670Fabjan,2014MNRAS.442.2304Hirschmann,2015MNRAS.446..521Schaye}.
%dPN{Please try and include more general references from all the big numerical groups where possible - like Springel, Schaye, Frenk as well.}

\subsection{Feedback schemes and resolutions}

In this work we focus on  six Dianoga regions re-simulated with three models at low-resolution (1x), each model with different softening and feedback schemes. One of these models has also a high-resolution  counterpart  in order to isolate the effect of resolution alone (hereafter 10xB20 simulations, where ``10x'' means it has been re-simulated with $10$ times lower particle mass). The four suites are the following: 

\begin{itemize}
\item 1xR15: this model, described in \citetalias{2015ApJ...813L..17Rasia} and in \cite{2017MNRAS.467.3827Planelles},  uses the SMBH feedback scheme presented in \cite{2015MNRAS.448.1504Steinborn}, and implements both mechanical and radiative feedback.
The  model is parameterised by an outflow efficiency $\epsilon_o$ that  regulates gas heating power  $P_0$ used for mechanical feedback, while the radiative efficiency $\epsilon_r$ regulates the  luminosity of the radiative component. The feedback energy per unit time in this model is then the sum of $P_0$ and the fraction $\epsilon_f$ of the luminosity (see Equations 7-12 in \citealt{2015MNRAS.448.1504Steinborn} for more details).
This model is an improvement on the model presented in \cite{2014MNRAS.442.2304Hirschmann}, wherein a constant radiative efficiency is assumed that does not allow for a smooth transition between the quasar-mode
and the radio-mode. In the model presented in \cite{2015MNRAS.448.1504Steinborn} the amount of energy released by SMBH growth is determined by the radiative efficiency factor $\epsilon_r$ and the fraction that is thermally coupled with the surrounding gas is denoted by $\epsilon_f$. These simulations have been used in \citetalias{2020Sci...369.1347Meneghetti} in order to compare simulations with observational data. \citetalias{2015ApJ...813L..17Rasia} showed that their model produces cool-core clusters in similar proportions to observations from \cite{2011A&A...526A..79Eckert}.
\item 1xRF18: this model is described in \cite{2013MNRAS.436.1750RagoneFigueroa} with some modifications introduced in \citetalias{2018MNRAS.479.1125RagoneFigueroa}. 
%and , and is based on \cite{2005MNRAS.361..776Springel}.
%The main difference with the model of \cite{2005MNRAS.361..776Springel} is that gas accreted onto BHs is not removed from the surrounding gaseous component. The small price for a small mass non-conservation (negligible on galactic scales), avoids the gas depletion in the BH surroundings and also shallows out the inner part of the cluster gravitational potential, which   makes it easier for BHs to drift away. This effect is counter-acted by a re-positioning scheme applied at each time-step that sets the BH position to the one of the nearest particle with minimum gravitational potential within its softening length. 
%The energy from AGN is not simply added to gas particles as in the original model from \cite{2005MNRAS.361..776Springel}, as star-forming particles would lose this contribution immediately due to the multi-phase model from \cite{2003MNRAS.339..289Springel}. For this reason \citetalias{2018MNRAS.479.1125RagoneFigueroa} 
At variance with \cite{2005MNRAS.361..776Springel}, \cite{2013MNRAS.436.1750RagoneFigueroa} implemented a mechanism of cold-cloud evaporation, so that if gas in cold phase is heated by the AGN energy to a temperature that exceeds the average gas temperature, then the corresponding particle is removed from the multi-phase state to avoid star formation. To prevent the particle from re-entering the multi-phase state in the next time-step, they add a maximum temperature condition for a particle to be star-forming \citep[as explained in Sec. 2.1 of][]{2013MNRAS.436.1750RagoneFigueroa}.
These simulations have a larger softening compared to 1xR15.  
The simulated BCG mass evolution and the BCG alignment with the host cluster are in agreement with observational data \citep{2018MNRAS.479.1125RagoneFigueroa,2020MNRAS.495.2436RagoneFigueroa}. 
%from \cite{2018AstL...44....8Kravtsov}.
\item 10xB20: this model  is presented in \cite{2020A&A...642A..37Bassini} (B20), where the regions are re-simulated with a mass resolution $10$ times higher than the above-mentioned 1x models. They have a feedback scheme similar to 1xRF18, however  they do not implement cold cloud evaporation in order to reduce the feedback efficiency. As a consequence, the sample reproduces the stellar mass function (SMF) from \cite{2013MNRAS.436..697Bernardi} over more than one order of magnitude in the intermediate SH mass regime, however it overproduces the number count of most massive galaxies.
\item 1xB20: we created this new sample to isolate the effect of resolution on subhaloes. We re-simulated four regions with the same feedback configuration  of \citetalias{2020A&A...642A..37Bassini} 10xB20 and re-scaled their softening parameters to the 1x resolution level. 
\end{itemize}
  
  In Table \ref{tbl:ps} we report all feedback parameters, softening and mass resolution values for the four suites 1xR15, 1xRF18, 10xB20, and 1xB20.
  There we report the minimum gas temperature allowed for cooling, min $T_g,$  an outflow efficiency $\epsilon_o,$ the BH radiation efficiency $\epsilon_r,$ and the feedback efficiency $\epsilon_f$ \citep[see][for more details]{1973A&A....24..337Shakura,2015MNRAS.448.1504Steinborn}. These parameters have been been tuned to fit some observational properties of galaxy
clusters best:
  %to best fit different observational properties of galaxy clusters: 
  \citetalias{2015ApJ...813L..17Rasia} and \citetalias{2020A&A...642A..37Bassini} tuned the  stellar mass $vs.$ black hole (BH) mass  relation $M_\star-M_{\rm BH},$ i.e. the Magorrian relation \citep{1998AJ....115.2285Magorrian}, and the stellar mass function (in the case of 10x), while \citetalias{2018MNRAS.479.1125RagoneFigueroa} tuned the Magorrian and the BCG mass $vs.$ $M_{\rm 500c}$ relation.

Note that even if some re-simulations have similar feedback parameter values, they have different feedback implementations (see discussion in Sec. 2 and Sec. 3 of \citetalias{2020A&A...642A..37Bassini}).
In particular the \citetalias{2018MNRAS.479.1125RagoneFigueroa} scheme takes into account cold cloud evaporation%~\citep[as described in][]{2010MNRAS.405.1491Murante} 
~\citep[see Appendix in][]{2013MNRAS.436.1750RagoneFigueroa}
and uses a  multiphase particle criterion depending not only on density but also on temperature. On the other hand,
the \citetalias{2020A&A...642A..37Bassini} scheme removes 
the temperature
criterion. Thus, the high density particles cool more efficiently ~\citep[see Appendix in][]{2013MNRAS.436.1750RagoneFigueroa}.
As a result, the \citetalias{2020A&A...642A..37Bassini} scheme leads to a better agreement of SMF with observations and   has  less efficient feedback than  to 1xRF18.
Unlike the other three setups, the 1xB20 realization has not be tuned with respect to any observational data.

\subsection{Zoom-in regions}
\label{sec:samples}

First off, in the following analyses, we focus primarily on SHs with $M_\mathrm{\rm SH}>2\times10^{10}\;h^{-1}{\rm M}_\odot$, whose inner structure is resolved with $\gtrsim100$ particles. 
In each region we focus on the main central halo at $z=0.4,$ in order to consistently compare the redshift of observations presented in \cite{2019A&A...631A.130Bergamini}  and \cite{2022A&A...659A..24Granata}. We consider three orthogonal lines-of-sight to each cluster halo, and extract all SHs in cylinders with depth $10$ comoving Mpc$/h$ and radius $R<0.15R_{\rm vir}$ centred on the halo centre. This radius is roughly consistent with that of the region that typically contains the cluster critical lines for strong lensing (see e.g. \citetalias{2020Sci...369.1347Meneghetti}).  

In  Table \ref{tbl:ds} we present properties of the six regions (D1, D3, D6, D10, D18, D25) used in this work, all with a virial mass $M_{\rm vir}>3\times10^{14}\;h^{-1}{\rm M}_\odot.$ 
All regions that have been re-simulated assume a $\Lambda$CDM cosmology with parameters $\Omega_m=0.24,\Omega_b= 0.04,$ $n_s= 0.96,\sigma_8 = 0.8,$ $h= 0.72.$ 

 Table \ref{tbl:ds} shows virial masses and the number (averaged over the three different projections) of SHs of the $6$ regions in the four suites. From the table it is already possible to appreciate a difference in the amount of SHs identified in different simulations, with smaller softening and particle mass values generally leading to the formation of a larger number of SHs.

\subsection{SH selection methods}
 \label{sec:select}

 The haloes and SHs are identified using the FoF halo finder~\citep{1985ApJ...292..371Davis,2005MNRAS.364.1105Springel} and an improved version of the SH finder SUBFIND \citep{2001MNRAS.328..726Springel}, respectively. The latter takes into account the presence of baryons \citep{2009MNRAS.399..497Dolag}.
 
 For each region we identify the most massive FoF halo and its centre of mass,  SH particles are found by SubFind that iteratively removes unbound particles within the contour that traverses the gravitational potential saddle points~\citep[see][for more information on its accuracy]{2011MNRAS.410.2617Muldrew}.
 
 In this work we will consider stellar masses hosted in SHs as the sum of all bounded star particles within $50$ physical projected kpc: this aperture is chosen as the one in observations of \cite{2018AstL...44....8Kravtsov}.

\begin{figure}
   \centering
    \includegraphics[width=0.9\linewidth]{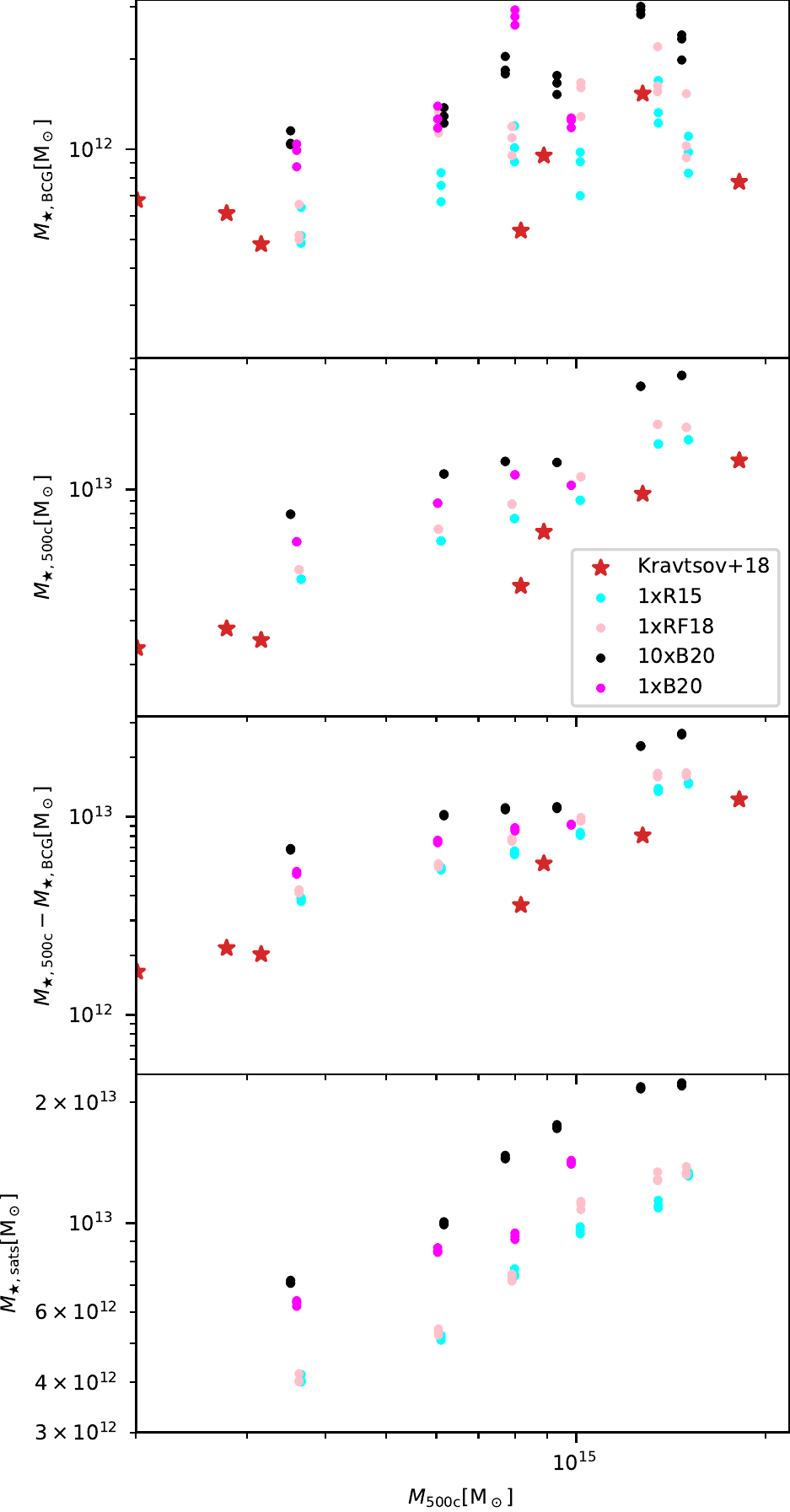}
   \caption{Comparison of the values of BCG mass (first panel), stellar mass  (second panel),  satellite (i.e. stellar mass minus BCG mass, third panel) and the  stellar mass sum of all well-resolved  SHs within $R_{\rm vir}$ (fourth panel)    within $R_{\rm 500c}$  for our four suites (circles coloured as in the label) and with the observational data from  \cite{2018AstL...44....8Kravtsov} (red stars). BCG masses are computed within a projected aperture of $50$ physical kpc and for each region we present masses from three orthogonal projections.
   From left to right, data points corresponds to the main haloes of the following regions: D25, D3, D6, D18, D10, and D1 presented in Table \ref{tbl:ds}.
   }
   \label{fig:kravtsov}%
\end{figure}

\section{Results}
\label{sec:discu}

\subsection{SH masses}

The top panel of Figure \ref{fig:kravtsov}  
shows the BCG stellar masses against the total mass within $R_{\rm 500c}$
of our four suites and observations from \cite{2018AstL...44....8Kravtsov}. Here we can see that 1xR15 and 1xRF18 have BCGs that tend to agree with observations, while \citetalias{2020A&A...642A..37Bassini}s 1x and 10x simulations have BCGs much brighter than observations, as expected with their low feedback.

We will now show that the overproduction of stars is not limited to BCGs: the total stellar mass within $R_{\rm 500c}$ (i.e. $M_{\star,\rm 500c},$ presented in second panel of Fig. \ref{fig:kravtsov}) is systematically higher than observations for both \citetalias{2020A&A...642A..37Bassini} models, while \citetalias{2018MNRAS.479.1125RagoneFigueroa} and \citetalias{2015ApJ...813L..17Rasia}  produce  a lower amount of stars, closer to the observed values.

In the third panel of Fig. \ref{fig:kravtsov} where we compare the stellar mass in satellites as estimated by \cite{2018AstL...44....8Kravtsov}, as the difference between  $M_{\star,\rm 500c}$ and the BCG stellar mass,    we see the same trend as for $M_{\star,\rm 500c}.$
Finally, to consistently compare simulations with different resolutions, and to rule out that this overproduction of stars is caused by  intracluster light (ICL)  instead of being a problem of SHs,  we compare total stellar masses of only  well-resolved satellites  (i.e. all sub-haloes with mass $M_{\rm SH}>2\times10^{10}{\rm M}_\odot,$ as defined in Sec. \ref{sec:select}).
In the bottom panel of Fig. \ref{fig:kravtsov} we show that the  \citetalias{2020A&A...642A..37Bassini} model produces the SHs with the highest stellar masses, followed by \citetalias{2015ApJ...813L..17Rasia}
 and \citetalias{2018MNRAS.479.1125RagoneFigueroa}. In addition, 10xB20 has systematically higher stellar masses than 1xB20, which shows that with increasing resolution there is an increase of high-stellar mass SHs.
 
\begin{figure*}
   \centering
    \includegraphics[width=0.9\linewidth]{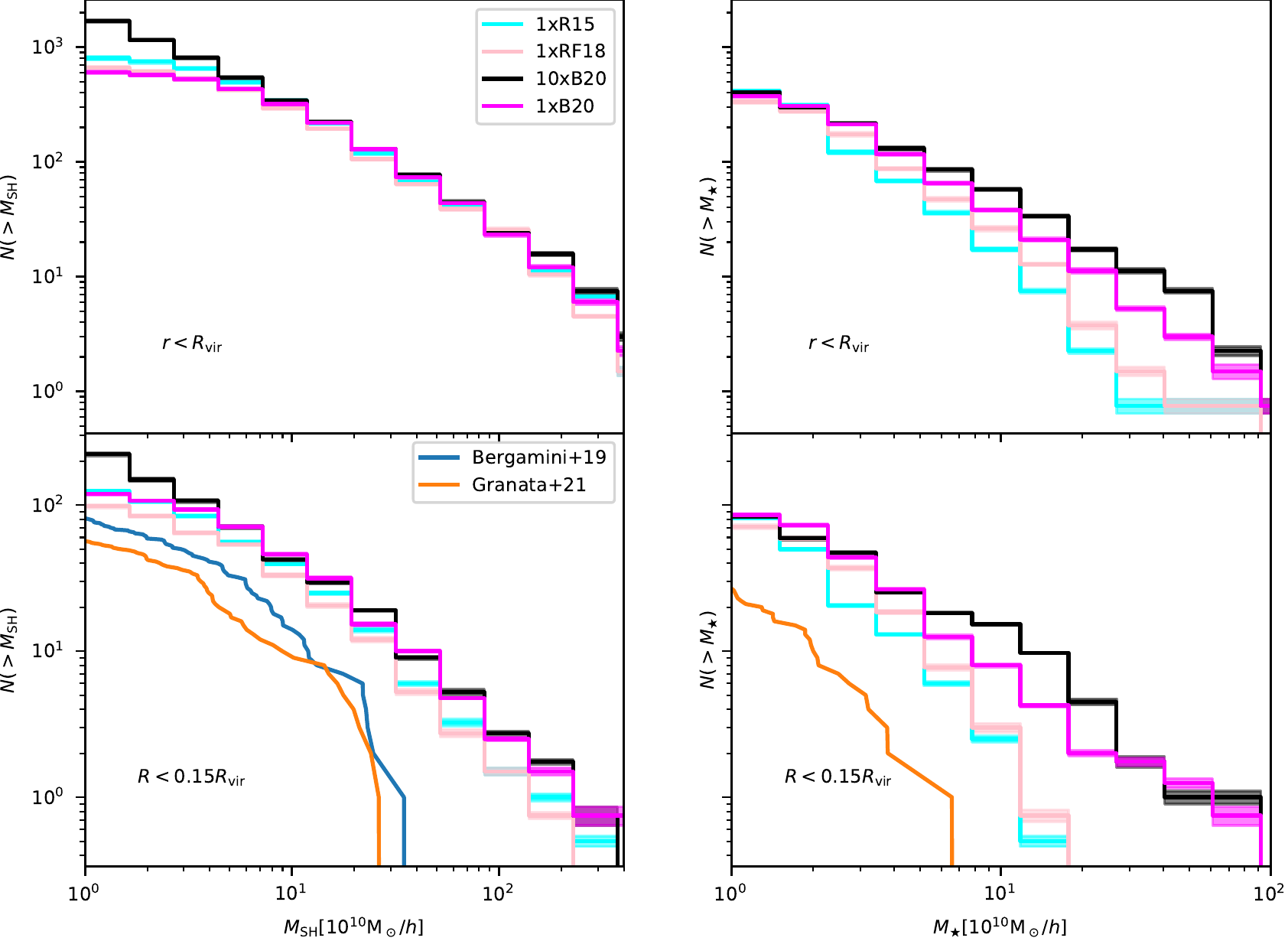}
   \caption{Average cumulative number of satellite substructures with total mass larger than $M_{\rm SH}$ and lower than $10^{13}{\rm M}_\odot/h$  (left column) and for stellar masses larger than $M_{\star},$ computed within an aperture of $30$ physical kpc.  We show data for the 1xR15 (cyan), 1xRF18 (pink),  10xB20 (black), and 1xB20 (magenta) four simulations in common between the four suites (i.e. D3, D6, D18, and D25 presented in Table \ref{tbl:ds}). For each simulation we average data over three orthogonal projections. Shaded area represent standard deviation between different setups. Upper panels consider SHs within the virial radius  and lower panels consider SHs with projected cluster-centric distance $R<0.15R_{\rm vir}.$ Left panels show cumulative number of total sub halo mass $M_{\rm SH},$ while right panels show cumulative mass of galaxy stellar mass associated to each SH. Observations are from central region ($R<0.15R_{\rm vir}$) of cluster AS1063 presented in \cite{2019A&A...631A.130Bergamini} and \cite{2022A&A...659A..24Granata}. Both simulations and observations have an average redshift of $z\approx 0.4.$}
   \label{fig:cmsub}%
\end{figure*}

In Fig.~\ref{fig:cmsub}, we show the average cumulative satellite SH mass number count (left column for total mass and right column for stellar mass)     both within the virial radius (upper panels) and within a cluster-centric distance $R<0.15R_{\rm vir}$ (bottom panels) for the  four regions (D3, D6, D18, and D25) presented in Table \ref{tbl:ds} that are in common between all suites. 
We also compare the theoretical estimates against the observed mass distributions derived by \cite{2022A&A...659A..24Granata} and  \cite{2019A&A...631A.130Bergamini} for the cluster AS1063.
%We also compare against SH total masses of observations from and  of the galaxy cluster AS1063, where we see that simulations over-estimate SHs number counts.
%and we note that observations consider only sub-structures with central velocity dispersion $\sigma>80$ km $s^{-1}$ and mag F160W (AB)$<24,$ while we applied no cut in simulated SHs as it would only slightly affect low-mass SHs\footnote{
%We assume that a value of mag F160 (AB)$=24$ corresponds to a stellar mass $M_\star=4.5\times10^{8}M_\odot,$ see Sec. 3.3.1 in \cite{2015ApJ...800...38Grillo}.
%}. 

Independently of the simulation set-up, all SH total-mass  count functions within $R_{\rm vir}$ (Fig. \ref{fig:cmsub} top-left panel) resemble a power-law, 
in agreement with other theoretical studies~\citep{2008MNRAS.386.2135Giocoli}. 
The 1x simulations deviate from the power-law behaviour at masses $\lesssim 2\times 10^{10}\; h^{-1}{\rm M}_\odot$, indicating that resolution limits become significant at these mass scales. This result validates our choice of excluding from our analysis the SHs below the mass limit discussed in Sec. \ref{sec:samples}.

If we focus on the total-mass count  within $0.15R_{\rm vir},$ we find that different setups produce significantly different number of SHs, with larger softening leading to less SHs. This is probably due to the fact that   large-softening simulations produce more fragile SH cores which are less resistant to tidal forces.

We further study the over-production of stars by showing the cumulative number count of SH with a given stellar mass. We present the cumulative number within $R_{\rm vir}$ in top-right panel of Fig.~\ref{fig:cmsub}, where we see that 1xR15 and 1xRF8 have very similar stellar masses (with a small difference for the lowest massive SHs), while 1xB20 has much more bright galaxies (as expected by its lower feedback). 1xB20 and 10xB20 have nearly the same stellar mass counts.
The stellar mass count inside $0.15R_{\rm vir},$ (Fig. \ref{fig:cmsub} bottom-right panel) shows that a better resolution increases the number of high-stellar mass galaxies.

In particular,  \cite{2022A&A...659A..24Granata} use a
\cite{1955ApJ...121..161Salpeter} IMF \citep[see Appendix in][]{2021A&A...656A.147Mercurio}, whereas in our simulations we adopt a \cite{2003PASP..115..763Chabrier}. We therefore  re-scaled observational stellar masses by a factor $0.58$ to match values of a Chabrier IMF~\citep[as proposed in][]{2014ApJS..214...15Speagle}.

From the figure, we can evince that both feedback schemes and resolution parameters strongly affect the SH population in galaxy cluster cores, compared to their effect on the population within $R_{\rm vir},$ especially when it comes to high-mass SHs ($M_{\rm SH}>4\times10^{11}M_\odot/h$): there are more massive SHs that reach the core of galaxy cluster when increasing resolution, or lowering feedback.
We also note that 1xRF18 and 1xR15 simulations are the ones that best match observations of the galaxy mass distribution in the internal region of galaxy clusters.

\subsection{SH radial distributions}
\begin{figure}
   \centering
    \includegraphics[width=0.9\linewidth]{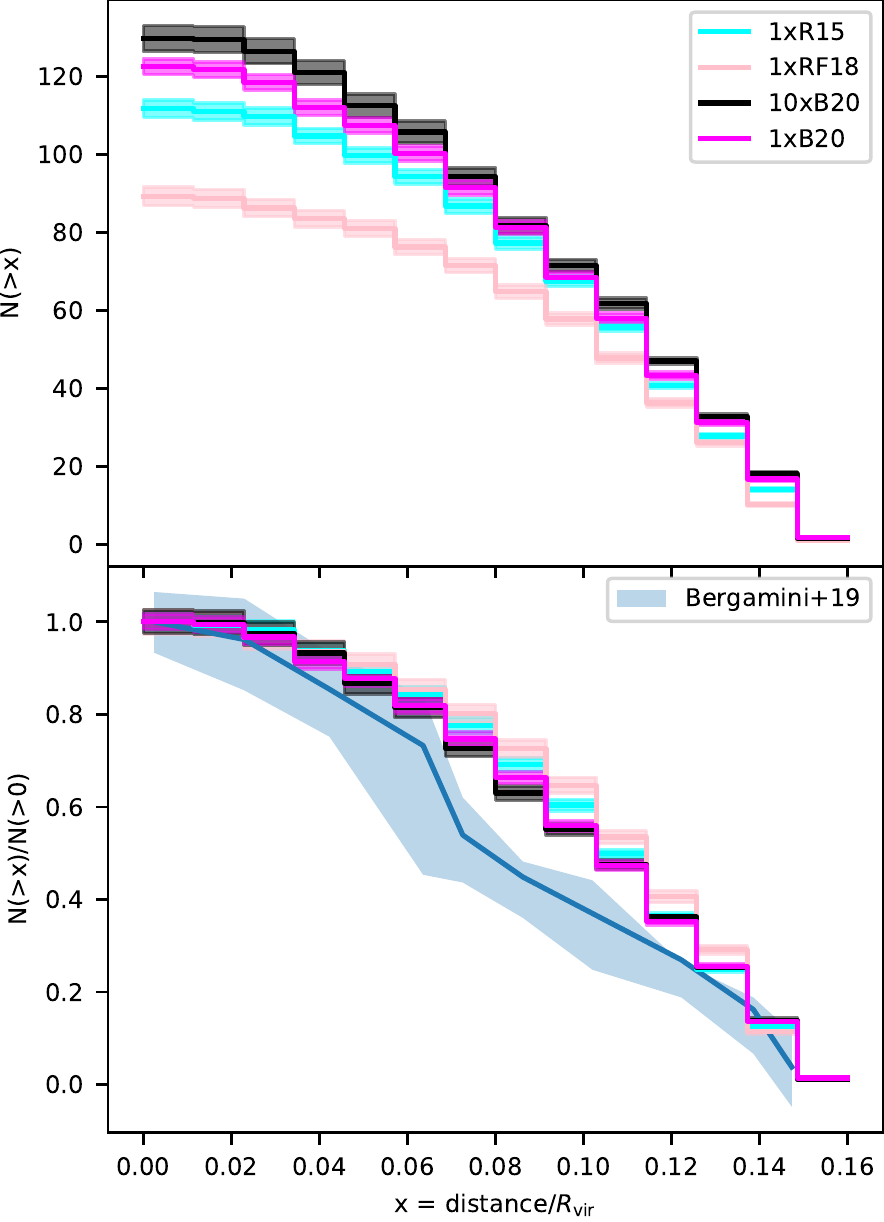}
   \caption{Average cumulative number of SHs (left panel) and the same value normalised to unity (right panel) within a given projected cluster-centric distance in units of the host virial radius. Here we only consider only the four Dianoga regions in common between all four setups in Table \ref{tbl:ds} and SHs within a projected distance $R<0.15R_{\rm vir}$ and with a total mass greater than $2\times10^{10}h^{-1}{\rm M}_\odot$. The black, cyan, and pink lines refer to the 10x, 1xR15, and 1xRF18 simulations, respectively. Shaded area shows data from \cite{2019A&A...631A.130Bergamini}.
   }
   \label{fig:m20-dx}%
\end{figure}

The impact of resolution and feedback on the total number of SHs in the central regions of clusters can be  better appreciated in Fig.~\ref{fig:m20-dx} (top panel), where we show the average number of SHs with projected cluster-centric distance larger than a given fraction of the host virial radius. This is shown for the four regions in common between our four setups, namely, D3, D6, D18, and D25 as in Table. \ref{tbl:ds}.

The average number of SHs in  central region of clusters in the 1xRF18 simulations is smaller than the corresponding number in the 1xR15 simulations by $\sim 25\%$ (see Table \ref{tbl:ds}). 
This behaviour is similar to what we found in Fig. \ref{fig:cmsub} (bottom-left), and is most probably caused by the relation between SH fragility in cluster cores and softening lengths.

From the bottom panel of Figure \ref{fig:m20-dx} where we show the normalised distributions, we notice that simulations have all a very similar spatial concentration of SHs.
If we compare the relative compactness with observations from  \cite{2019A&A...631A.130Bergamini} (as done already in \citetalias{2020Sci...369.1347Meneghetti} for 1xR15 simulations) we find that all suites are unable to reproduce the drop in number-count in the region   $[0.07-0.1]R_{\rm vir}$ that is found in observations.

\subsection{Circular velocity vs sub-halo mass}

\begin{figure*}
   \centering
    %\includegraphics[width=\textwidth]{ns_profile_vir_vir10.png}
    %$$\Delta_{\tt vir}$$
    %\includegraphics[width=\textwidth]{hod_vir_minitabella_ABs_bars}
    \includegraphics{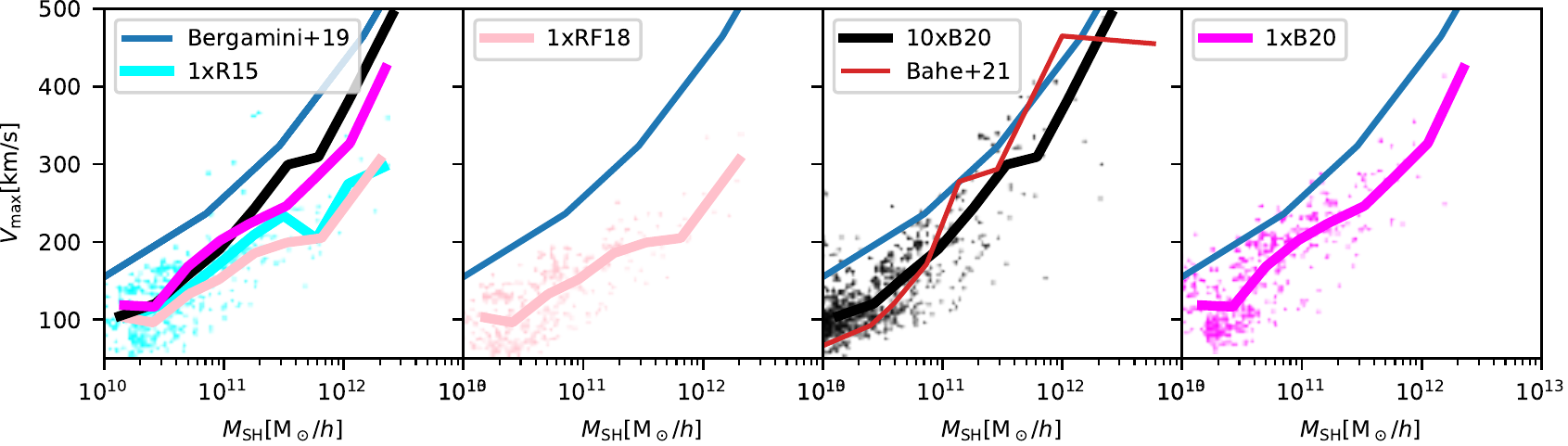}
   \caption{
   $V_{\rm max}$ 
   vs. SH mass for 10xB20 simulations (left panel), 1xR15 (central panel) and 1xRF18 (right panel), for all SHs (except for BCGs) with a projected distance ($R<0.15R_{\rm vir}$).  Solid line colours are set as in Fig. \ref{fig:logslopes-xcomp-2e11-4e11}. Over-plotted in blue the relation from \cite{2019A&A...631A.130Bergamini}. We report data for  10xB20 simulations (black solid line), 1xR15 simulations subset of 10xB20 regions (solid cyan line), 1xRF18 simulations subset of 10xB20 regions (pink line) and 1xf$_{\rm SH}$ (magenta line), for SHs with projected distances $<0.15R_{\rm vir}$ (right panel). We compare our high-resolution simulations with simulated data points of high-resolution Hydrangea simulations \citep{2021arXiv210901489Bahe}  (red).}
   \label{fig:m20-sinistra}%
\end{figure*}

In Fig. \ref{fig:m20-sinistra} we show the SH distribution in the plane $V_{\rm max}-M_{\rm SH}$ for all simulation suites, where $V_{\rm max}$ is computed as in Eq.~\ref{eq:vcirc}. Each data-point corresponds to a different SH. 

The SHs in the 10xB20 and 1xB20 simulations generally have higher maximum circular velocities than in the 1xR15 and 1xRF18 simulations. The amount of this difference grows as a function of the SH mass. Consequently, the $V_{\rm max}-M_{\rm SH}$ relations in the 10xB20 and 1xB20 simulations are significantly steeper than the others.

For comparison, Fig. \ref{fig:m20-sinistra} shows  the $V_{\rm max}-M_{\rm SH}$ relation of \cite{2019A&A...631A.130Bergamini} (blue solid lines),  based on the strong lensing analysis of galaxy clusters MACSJ1206.2-0847 ($z=0.44)$, MACSJ0416.1-2403 ($z=0.4$), and AbellS1063 ($z=0.35$). \citetalias{2020Sci...369.1347Meneghetti} already showed that the SHs in the 1xR15 simulations fell below the observational relation of \cite{2019A&A...631A.130Bergamini} at all masses. Our results show that the same result holds for the 1xRF18 simulations. On the other hand, the gap between observations and simulations is reduced when the AGN feedback is less efficient, i.e. in the 10xB20 and 1xB20 simulations, although only for the SHs with the largest masses. 

This behaviour is akin to the one reported by \cite{2021MNRAS.505.1458Bahe}, who found that the SHs in the Hydrangea cluster simulations, implementing the  Eagle galaxy formation model \citep{2016MNRAS.456.1115Bahe}, have $V_{\rm max}$ similar or even exceeding the \cite{2019A&A...631A.130Bergamini} relation at masses $M_{\rm SH}\gtrsim 10^{11}\;h^{-1}{\rm M}_\odot$. Thus, we may interpret their results as an indication that in the sub-grid model implemented in Hydrangea SF is less efficient at a fixed halo mass.
%than in our 10xB20 and 1xB20 simulations.

At lower masses ($M_{\rm SH}\approx2\times10^{10}M_{\rm}$), the median SH maximum circular velocity in all our simulations reaches a value of $\sim 100\,\mathrm{km}\,s^{-1}$, almost independently of the resolution and feedback model. This value is very similar to the median value reported by \cite{2021MNRAS.505.1458Bahe} in the same SH mass-range, and is significantly below the observational relation of \cite{2019A&A...631A.130Bergamini}. For example, at masses of $\sim 10^{10}\,h^{-1}{\rm M}_\odot$, \cite{2019A&A...631A.130Bergamini} report an average $V_{\rm max}$ of $\sim 170\,\mathrm{km}\,s^{-1}$. In a more recent work focused on AS1063 and implementing a different approach to model the contribution of the cluster galaxies to the cluster strong lensing signal, \cite{2022A&A...659A..24Granata}  found that the typical maximum circular velocity of SHs in this mass-range is $\sim 150\,\mathrm{km}\,s^{-1}$.

We note that this same trend holds for the high-resolution simulations presented in \cite{2021MNRAS.505.1458Bahe}.
\cite{2021MNRAS.505.1458Bahe} reported the existence of a second branch of the $V_{\rm max}-M_{\rm SH}$ relation, followed by a minority of SHs with masses $M_{\rm SH}\lesssim 10^{11}\;h^{-1}{\rm M}_\odot$, that is consistent with observations. Even in our simulations with the highest mass resolution, we do not find evidence for a bimodal $V_{\rm max}-M_{\rm SH}$ relation, although we notice that few s  SHs in the 10xB20 and 1xB20 simulations have $V_{\rm max}$ close to and even higher than the \cite{2019A&A...631A.130Bergamini} relation.

For the above reasons, and to fairly compare with the data, in the next analyses we will divide SHs in two samples; low mass SHs with $M_{\rm SH}\in[4,6]\times10^{10}{\rm M_\odot}/h$ in order to have a large sample of both well-resolved sub haloes on a narrow mass-range; and high mass SHs with  $M_{\rm SH}\in[1,6]\times10^{11}{\rm M_\odot}/h,$ where the range is large enough to have a representative number of SHs in the high mass regime. 
The low mass SHs sample will help us study the mass-range observed in \cite{2019A&A...631A.130Bergamini}, while the high mass SHs sample will help study the mass-range where AGN feedback is most effective.

\subsection{Mass and circular velocity profiles}

 \begin{figure*}
   \centering
    \includegraphics{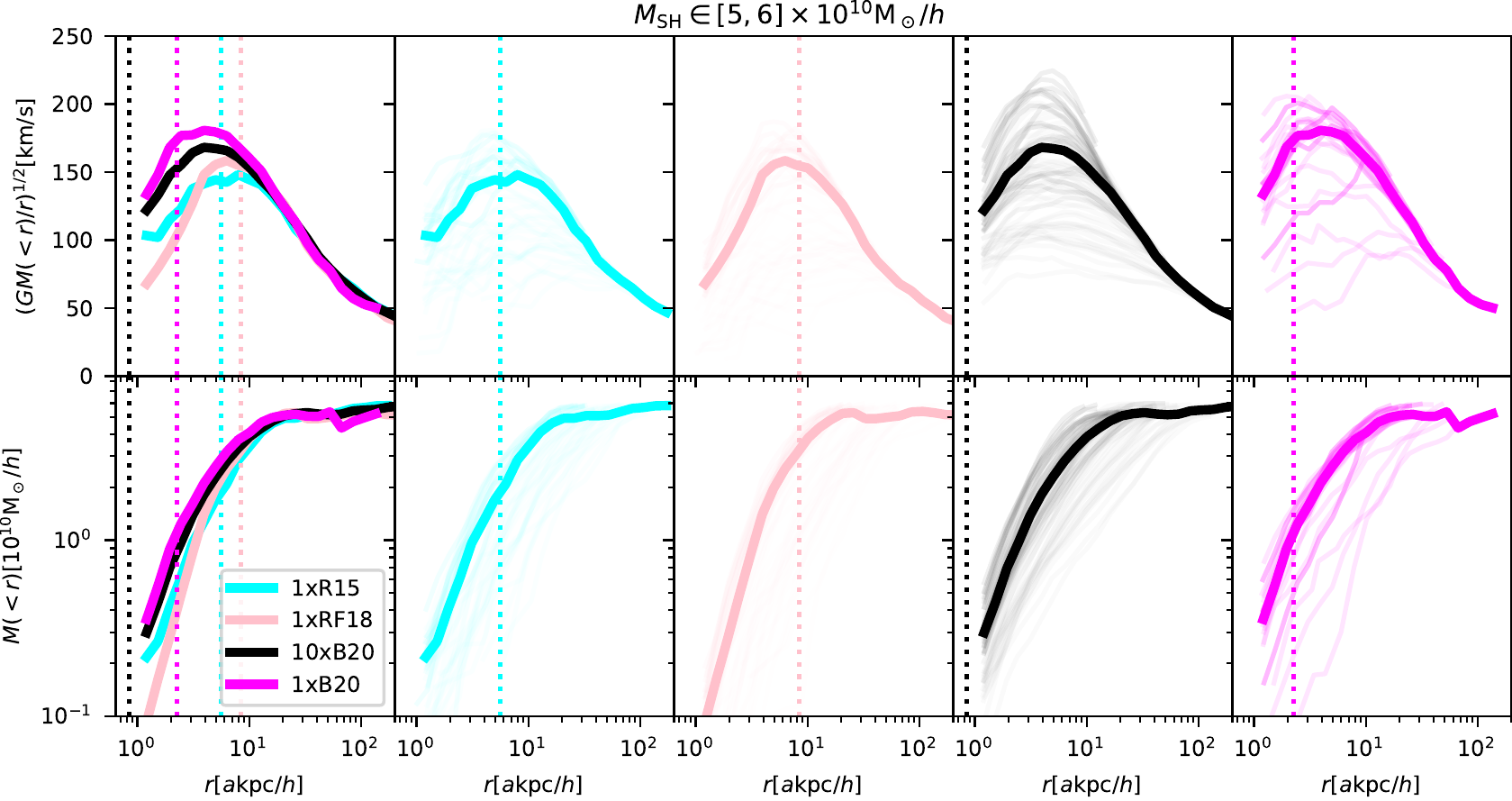}
   \caption{ Circular velocity profiles from total matter content of the SHs in the four cluster samples (upper row) and cumulative mass (lower row). The left most column shows the median values (thick solid lines) in radial bins for the 1xR15, 1xRF18, 10x, and 1xB20 simulations.
   The panels in other columns show profiles separately for each setup and single SHs (thin solid lines) together with their median values.    SHs mass-range is $M_{\rm SH}\in\left[5   \times10^{10},6\times10^{10}\right]{\rm M}_\odot/h,$ with a projected distance smaller than $0.15R_{\rm vir}.$   For each setup we plot    
   the   fiducial stellar resolution limit 
   $2.8\epsilon_{\star}$ as vertical dashed lines.
}
   \label{fig:V-M-lowm}%
\end{figure*}

\subsubsection{Low mass SHs}

We turn our attention to the SH inner structure, which we quantify by means of the SH mass and circular velocity profiles.

As shown in Fig.~\ref{fig:V-M-lowm}, the average difference between maximum circular velocities of SHs with masses $5\times 10^{10}\;h^{-1}{\rm M}_\odot \leq M_{\rm SH} \leq 10^{11}\;h^{-1}{\rm M}_\odot$ in different simulations reduces to $\sim 15\%$ and becomes negligible at the lowest masses. 

We stress that this is the mass-range that is most relevant in galaxy-galaxy strong lensing (GSSL), and thus it seems that simulations run with different feedback schemes (including the one used in the Hydrangea simulations, as shown in Fig. \ref{fig:m20-sinistra}), resolution parameters, and softening lengths are unable to match the observed compactness.

\subsubsection{High mass SHs}

\begin{figure*}
   \centering
    \includegraphics{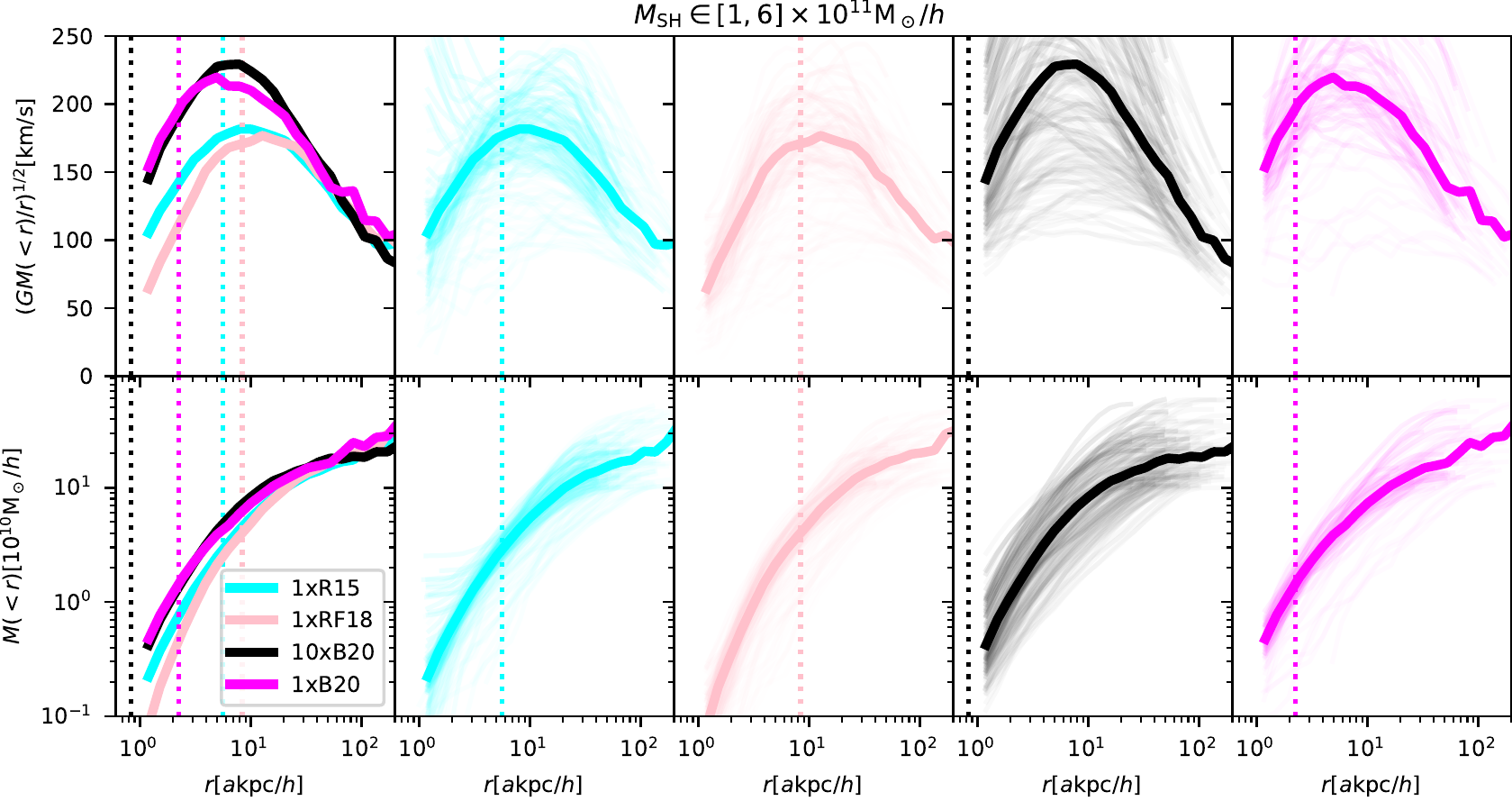}
   \caption{  Same as Fig. \ref{fig:V-M-lowm} but for  SHs with mass  $M_{\rm SH}\in\left[1   \times10^{11},6\times10^{11}\right]{\rm M}_\odot/h.$
   }
   \label{fig:V-M-him}%
\end{figure*}

The total SH circular velocity and mass profiles for all the four simulation suites are shown in  Fig.~\ref{fig:V-M-him}, which refers   to SHs with $M_{\rm SH}>10^{11}\;h^{-1}{\rm M}_\odot$. We notice that several SHs have maximum circular velocities that exceed $300\;\mathrm{km}\;s^{-1}$ in both the 10xB20 and 1xB20 simulations. The mean maximum circular velocity of these SHs is $\sim  220\;\mathrm{km}\;s^{-1}$. On average, the $V_{\rm max}$ of SHs with similar masses in the 1xR15 and 1xRF18 simulations are $\sim 20-30\%$ smaller.  

The SH mass and circular velocity profiles have limited dependence on the mass resolution. Indeed, on average, we cannot appreciate significant differences  between the profiles of SHs in the 1xB20 and 10xB20 simulations.

\begin{figure}
   \centering
    \includegraphics{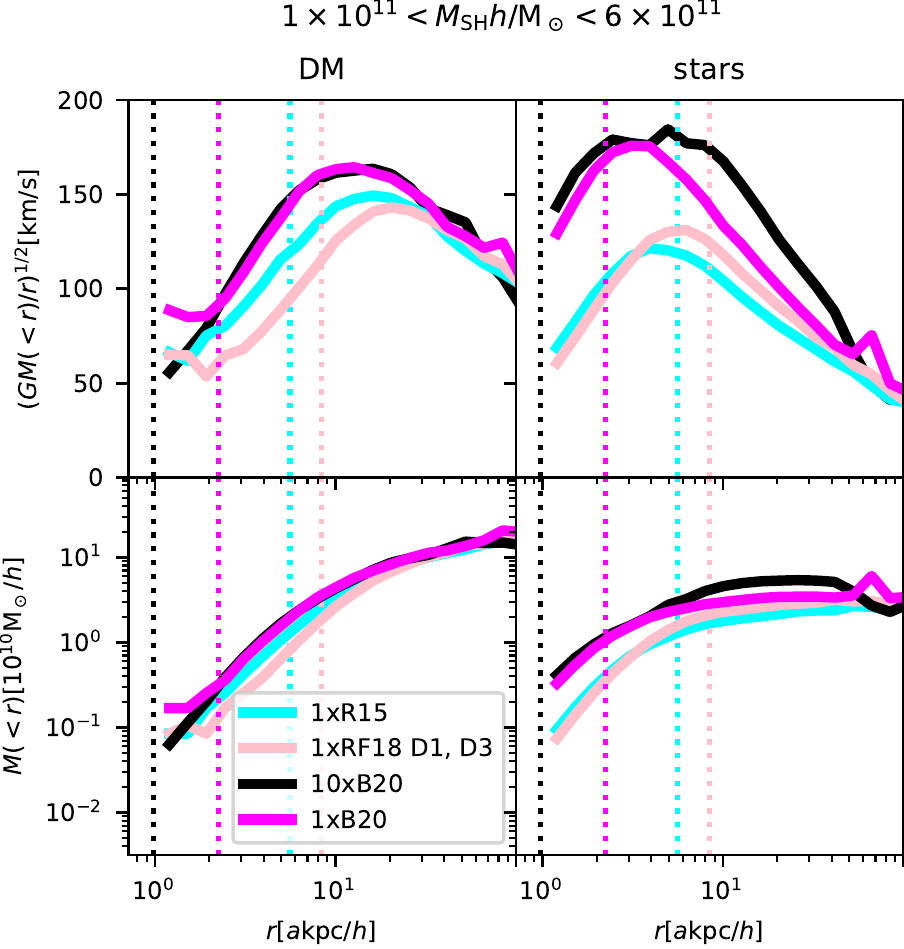}
   \caption{Average profiles of circular velocity  and cumulative mass per particle types: DM (left column), and stars (right column) for SHs with total mass in the range $M_{\rm SH}\in\left[1    \times10^{11},6\times10^{11}\right]{\rm M}_\odot/h,$ for the four simulated suites coloured as in Fig. \ref{fig:m20-sinistra}.  For each setup we plot 
   the corresponding 
   $2.8\epsilon_{\star}$ (where $\epsilon_\star$ is the stellar softening) as vertical dashed lines. Overall order of line colours from top to bottom is, respectively: black, magenta, cyan, pink. 
   }
   \label{fig:logslopes-xcomp-2e11-4e11}%
\end{figure}

The details of the DM and baryon mass distribution in the SHs have a significant dependence on the efficiency of the AGN energy feedback. The simulations with a less efficient feedback model (e.g. 10xB20 and 1xB20) cool more gas in the centre of their SHs. This process should lead to a more intense star formation in these regions, and the creation of dense stellar cores.  This effect is particularly evident in the SHs with large masses, as shown in the bottom panels of Fig.~\ref{fig:logslopes-xcomp-2e11-4e11}, where the average mass profiles and profiles of circular velocity of the DM and stars in SHs with mass $M_{\rm SH}>10^{11}\;h^{-1}{\rm M}_\odot$ are shown separately. In the 10xB20 and 1xB20 simulations, the stars dominate the total mass (i.e. sum of dark matter, gas, and star masses) profile within the inner $\lesssim 20-30\;h^{-1} $ comoving kpc.

The high central stellar density in these simulations should also trigger the  contraction of the dark matter haloes.
 Thus, the massive SHs in the 10xB20 and 1xB20 simulations are more compact and have higher maximum circular velocities compared to similar SHs in the 1xR15 and 1xRF18 samples. This behaviour is clear in the upper panels of Fig.~\ref{fig:logslopes-xcomp-2e11-4e11}, showing the circular velocity profiles of DM and stars in the massive SHs separately. The 1xR15 and 1xRF18 simulations differ significantly only in the very inner regions of the SHs, because of their different softening scales.

\begin{figure}
   \centering
     \includegraphics[width=0.9\linewidth]{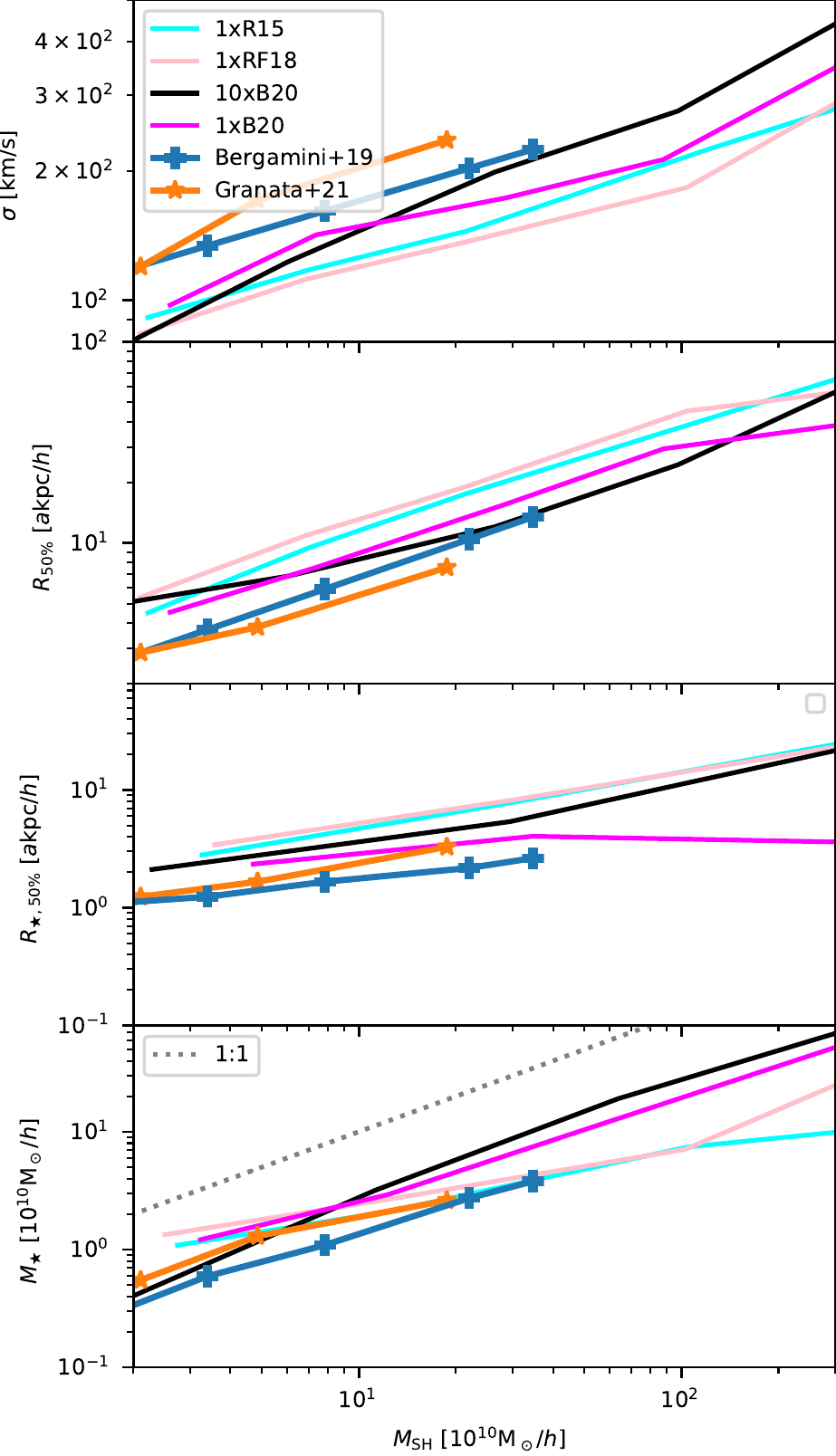}
   \caption{Median velocity dispersion $\sigma = V_{\rm max}/\sqrt{2}$ (top panel), median half total mass radius $R_{50\%}$ (second  panel), median half stellar mass projected radius $R_{\star,50\%}$ (third  panel),  median total stellar mass $M_\star$ (fourth and last panel) $vs.$ the total sub halo mass $M_{\rm SH}$ for satellites SHs.
   %We considered only  satellite SHs with $\sigma>80$km/s as in observations. 
   Simulation suites are coloured as in Fig. \ref{fig:cmsub}. We also show median values of  observations  from cluster AS1063 presented in \cite{2019A&A...631A.130Bergamini} (blue solid lines with crosses) and \cite{2022A&A...659A..24Granata} (orange solid lines with diamonds). We also show \cite{2022A&A...659A..24Granata} radii (third panel) and stellar masses (fourth panel) over \cite{2019A&A...631A.130Bergamini} SH masses in blue lines. Low-mass simulated SHs have stellar masses in agreement with observations but they are too large, while high-mass simulated SHs  are too numerous (no observations for SHs with $M_{\rm SH}>4\times10^{11}{\rm M}_\odot/h.$
   }
   \label{fig:mtot-vs-mstar}%
\end{figure}

\subsection{Comparison with Observations}

A similar analysis of the mass and velocity profiles of lower mass SHs indicates that the impact of AGN feedback and the differences between the simulation types are mass dependent. In   Fig.~\ref{fig:mtot-vs-mstar} we show the satellite SH central velocity dispersion $\sigma$ (first panel, defined as $V_{\rm max}/\sqrt{2}$ as in Appendix C of \citealt{2019A&A...631A.130Bergamini}),  the half mass radii (second and third panel, with $R_{50\%}$ for the total mass content and the projected stellar mass radius $R_{\star, 50\%}$) and the stellar mass $M_\star, $  (fourth panel) for SHs with velocity dispersion. %$\sigma>80$ km s$^{-1}$ as used in \citet{2019A&A...631A.130Bergamini}.
%Note that we consider only sub-structures with central velocity dispersion %$\sigma>80$ km $s^{-1}$ in agreement with observations, while observations have a also a cut on magnitude mag F160 (AB) = $24,$ which correspond to $\approx4.5\times 10^{8}M_\odot$~\cite{2015ApJ...800...38Grillo} and should not affect the results.

The first panel of Fig. \ref{fig:mtot-vs-mstar} shows qualitatively the same results as Fig. \ref{fig:m20-sinistra}. Here \cite{2019A&A...631A.130Bergamini} use a fixed $\sigma-M_{\rm SH}$ relation and determine the $\sigma-M_{\rm SH}$ obtained using using a Faber-Jackson relation whose normalisation and slope are constrained by the observed kinematics of cluster members.   \cite{2022A&A...659A..24Granata} has a fixed $R_{\star,50\%}-R_{\rm SH,50\%}$ relation and 
 use a Fundamental Plane (FP) relation, calibrated  using Hubble Frontier Fields photometry and data from the Multi Unit Spectroscopic Explorer on the
Very Large Telescope. The FP relation is then adopted to completely fix the velocity dispersion of all members from their magnitudes and half-luminosity and radii.
The second panel of  Fig. \ref{fig:mtot-vs-mstar} shows the half  total mass radius of SHs,   we see that simulations over-estimate the SH sizes. 
These values are  consistent with   their compactness over-estimation shown in the top panel of the same figure: in fact the maximum circular velocity is a proxy for SH compactness \citep{2019A&A...631A.130Bergamini}. 
We see the same behaviour in the half stellar mass radius presented in the third panel of   Figure \ref{fig:mtot-vs-mstar}, where we show the half-mass radii of the stellar component $vs.$ their SH mass. Here we compare our data with results from  \cite{2022A&A...659A..24Granata} using both their SH mass estimate (orange crosses) and the mass estimates from  \cite{2019A&A...631A.130Bergamini} (green diamonds), and we see that simulations do under-estimate stellar component compactness. 
In Fig. \ref{fig:mtot-vs-mstar} fourth panel we compare the   stellar mass against total mass of simulated SHs with observations, where we see that the stellar masses from low-mass simulated SHs ($M_{\rm}<\times10^{11}$) do overlap the values from observations. We therefore conclude that low-mass simulated SHs have   correct stellar fractions but a too-low compactness (see first and second panels of the Figure).

For what concerns high-mass SHs ($M_{\rm}>4\times10^{11}$ as in fourth panel of  Fig. \ref{fig:mtot-vs-mstar}), simulations do produce too many SHs: in fact,  observations have no SHs in this mass-regime.

\begin{figure}
   \centering
     \includegraphics[width=0.9\linewidth]{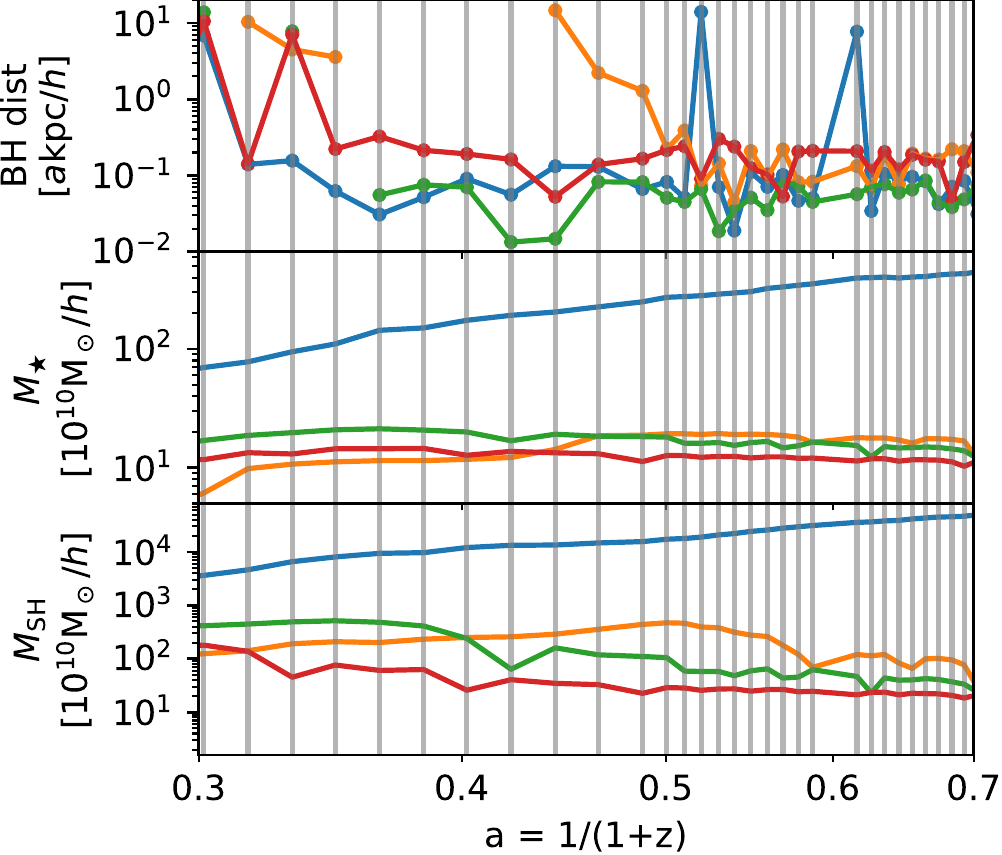}
   \caption{We track the 4 most massive central SHs of D6 10xB20. Upper panel shows the distance from the nearest most massive black hole within $20 a{\rm kpc}/h,$ central panel shows the evolution of the stellar mass and bottom panel shows the evolution of the total SH mass. Each colour represents a different SHs. The blue upper line is relative to the BCG.}
   \label{fig:bhdist}%
\end{figure}

 We finally ruled out the possibility that such an increase of stellar mass is due to a wandering BH particle. In Figure \ref{fig:bhdist} we track the four most massive SHs of the D6 (10xB20) back in time and show that they had a nearest BH (searched within a sphere of $20 a{\rm kpc}/h)$ and found that up to redshift $a\approx0.3-0.4$ three of the four SHs have a BH near the centre (with distance $<1$~kpc) and all of them have a stellar mass that is growing smoothly.
 
\begin{figure*}
   \centering
    %\includegraphics[width=\textwidth]{ns_profile_vir_vir10.png}
    %$$\Delta_{\tt vir}$$
    %\includegraphics[width=\textwidth]{hod_vir_minitabella_ABs_bars}
    \includegraphics{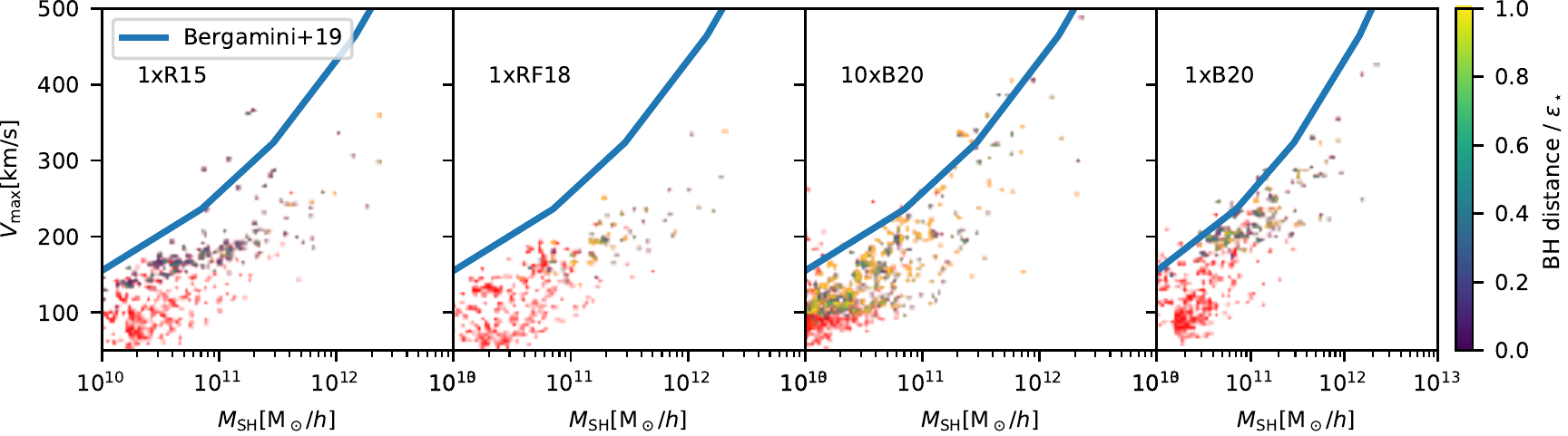}
   \caption{
   Same as Fig. \ref{fig:m20-sinistra}, where points are colour-coded (bar on the right) by the  distance of the nearest BH from the SH centre, in units of the respective gravitational softening length reported in Table \ref{tbl:ps}. Red points indicates SHs with no BH within $20$ comoving kpc$/h.$  }
   \label{fig:m20-sinistra-colour-bhd}%
\end{figure*}

 Additionally, in Fig. \ref{fig:m20-sinistra-colour-bhd} we show the $V_{\rm max}-M_{\rm SH}$ relation and colour-code SH points by the distance of the nearest BH distance in units of the softening lengths (reported in Table \ref{tbl:ps}).
 We notice that most of the massive haloes have a well-centred BH (i.e. within a gravitational softening radius), while low-massive ($M_{\rm}<10^{11}$) SHs tend to have no BHs. However in this mass-range the SHs without BHs have probably not been seeded yet, and  the $V_{\rm max}$  parameter is under control.

\section{Conclusions}
\label{sec:conclu}

We studied in detail the discrepancy between observations and simulations found in \citetalias{2020Sci...369.1347Meneghetti}, where they found that observations show much more compact SHs than simulations.
To this end we analysed the properties of SHs in the cluster core of several hydrodynamic   cosmological zoom-in simulations that were run with different resolution, feedback scheme and softening lengths.

In particular we studied $6$ Dianoga zoomed-in regions  re-simulated with: the fiducial model (1xR15) used in \citetalias{2020Sci...369.1347Meneghetti}; a model  with a larger softening and a different feedback scheme (1xRF18) and  one with  higher resolution (10xB20) where feedback parameters have been re-calibrated to match observations at that resolution.
We also run their 1x counterpart, 1xB20, with the same exact feedback scheme as 10xB20 simulations, in order to disentangle resolution effects from those related to different implementations of AGN feedback.

We found that:

\begin{itemize}
    \item Varying resolution level, softening lengths, and  feedback schemes  does not impact significantly the $V_{\rm max}-M_{\rm SH}$ relation   in the SH mass-range of interest for GSSL (i.e. $M_{\rm SH}<10^{11}{\rm M}_\odot/h$), and all the simulations that we considered are unable to reproduce SH compactness as observed by \cite{2019A&A...631A.130Bergamini}.  These same results holds for  Hydrangea simulations presented in \cite{2021MNRAS.505.1458Bahe}.
\item Some setups (1xB20, 10xB20 and Hydrangea simulations) are capable of producing a significant increase of  $V_{\rm max}$ in high mass SHs ($M_{\rm SH}>4\times10^{11}{\rm M}_\odot/h$), and they do match the observed $V_{\rm max}-M_{\rm SH}$ scaling relation in this mass-range.
However, we also found that the cause of their high $V_{\rm max}$  is  because these SHs have high and unrealistic stellar masses.
%~\citep[see comparison with][ in Fig. \ref{fig:kravtsov}]{2018AstL...44....8Kravtsov}. 
The fact that some of the current simulations produce exceedingly  high stellar masses was already found in works as \citetalias{2018MNRAS.479.1125RagoneFigueroa} %\cite{2020MNRAS.495.2436RagoneFigueroa}
and \citetalias{2020A&A...642A..37Bassini}, and most likely this very problem plagues the Hydrangea \citep{2021MNRAS.505.1458Bahe} simulations too.
\item Most importantly, observed galaxy cluster cores do not have as much high-mass ($M_{\rm SH}>4\times10^{11}{\rm M}_\odot/h$) SHs as the ones produced in simulations (see bottom-left panel in Fig. \ref{fig:cmsub}), thus the mass-regime where simulations are capable of matching observations  is not the one relevant in GSSL.
\item Given that different schemes produce different  $V_{\rm max}$ relations on high mass SHs, we find partial agreement with \cite{2021MNRAS.504L...7Robertson}, that current high-resolution simulations are not capable of constraining $\Lambda$CDM paradigm.
However, we find that this is true only for high-mass SHs ($M_{\rm SH}>4\times10^{11}{\rm M}_\odot/h$)  and that properties of low-mass SHs ($M_{\rm SH}<10^{10}{\rm M}_\odot/h$) can be used in future simulations to constrain  cosmology. 
\end{itemize}

In this work we  found  that the discrepancy between observations and simulations found in \citetalias{2020Sci...369.1347Meneghetti} cannot  be solved by simply calibrating the feedback efficiency of simulations, because this will lead to an unrealistically high number of bright galaxies, thus it seems that modern hydrodynamic simulations cannot reproduce both compactness and stellar masses of SHs in the internal regions of galaxy clusters.

Finally we found that the  $V_{\rm max}-M_{\rm SH}$ at $M_{\rm SH}\lesssim10^{11}{\rm M}_\odot/h,$ which is the most relevant mass-range in GSSL, is showing a tension between observations and simulations of different resolution and feedback parameters, and thus it is still challenging either the current feedback schemes or the underlying $\Lambda$CDM paradigm, or both.

\begin{acknowledgements}
We acknowledge support from the grant
PRIN-MIUR 2017 WSCC32. 
AR acknowledges support by  MIUR-DAAD contract number 34843  ``The Universe in a Box''. MV is supported by the Alexander von Humboldt Stiftung and the Carl Friedrich von Siemens Stiftung. MV and KD acknowledge support by the Deutsche Forschungsgemeinschaft (DFG, German  Research  Foundation)  under  Germany's  Excellence Strategy - EXC-2094 - 390783311. KD also acknowledges support through the COMPLEX project from the European Research Council (ERC) under the European Union’s Horizon 2020 research and innovation program grant agreement ERC-2019-AdG 882679. 
 We used  Trieste IT framework \citep{2020ASPC..527..307Taffoni,2020ASPC..527..303Bertocco}. 
We are especially grateful for the support by M. Petkova through the Computational centre for Particle and Astrophysics (C$^2$PAP). 
PN acknowledges the Black Hole Initiative (BHI) at Harvard University, which is supported by grants from the Gordon and Betty Moore Foundation and the John Templeton Foundation, for hosting her.
\end{acknowledgements}

% WARNING
%-------------------------------------------------------------------
% Please note that we have included the references to the file aa.dem in
% order to compile it, but we ask you to:
%
% - use BibTeX with the regular commands:
\bibliographystyle{aau} % style aa.bst
 \bibliography{bolo} % your references Yourfile.bib
%
% - join the .bib files when you upload your source files
%-------------------------------------------------------------------

\end{document}